\documentclass[12pt]{article}
\usepackage{graphicx}
\usepackage{latexsym}
\usepackage[algoruled]{algorithm2e}

\dontprintsemicolon
\SetKwFor{WhileDo}{while}{do begin}{end\rule[0cm]{0cm}{4.1mm}}
\SetKwFor{ForEachDo}{for each}{do begin}{end\rule[0cm]{0cm}{4.1mm}}

\newcommand{\tbf}[1]{\textbf{#1}}

\newcommand{\ttt}[1]{\texttt{#1}}
\newcommand{\tsc}[1]{\textsc{#1}}

\newcommand{\define}[1]{\emph{#1}}
\newcommand{\graph}[1]{\mathcal{#1}}
\newcommand{\network}[1]{\mathcal{#1}}
\newcommand{\vertices}[1]{\mathcal{#1}}
\newcommand{\edges}[1]{\mathcal{#1}}
\newcommand{\relation}[1]{\mathbf{#1}}

\newcommand{\between}[3]{#1 = #2,\ldots,#3}
\newcommand{\set}[1]{\{#1\}}
\newcommand{\edge}[2]{(#1\,{:}\,#2)}

\newcommand{\OO}{\mathcal{O}}
\newcommand{\RR}{\hbox{\sf I\kern-.14em\hbox{R}}}
\newcommand{\NN}{\hbox{\sf I\kern-.13em\hbox{N}}}
\newcommand{\HH}{\hbox{\sf I\kern-.13em\hbox{H}}}

\newtheorem{definition}{Definition}
\newtheorem{theorem}{Theorem}
\newenvironment{proof}{\noindent\tsc{Proof:} }{\strut~\hfill$\Box$}
\newenvironment{example}{\noindent\tbf{Example:} }{}
\newcommand{\url}[1]{\\\ttt{\tbf{\footnotesize{#1}}}}

\newcommand{\slika}[4][!ht]{
  \begin{figure}[#1]
    \centering
    \includegraphics[#2]{eps/#3.eps}
    \caption{\emph{#4}}
    \label{#3}
  \end{figure}
}

\newcommand{\slike}[4][!ht]{
  \begin{figure}[#1]
    \centering
    #2
    \caption{\emph{#4}}
    \label{#3}
  \end{figure}
}

\newcommand{\podslika}[3][]{
  \begin{tabular}{c}
    \includegraphics[#1]{eps/#2.eps} \\ #3
  \end{tabular}
}

\begin{document}


\title{Short Cycles Connectivity}

\author{
   Vladimir Batagelj and Matja\v{z} Zaver\v{s}nik \\
   University of Ljubljana, FMF, Department of Mathematics, \\
   and IMFM Ljubljana, Department of TCS, \\
   Jadranska 19, 1000 Ljubljana, Slovenia
}

\maketitle

\begin{abstract}
   Short cycles connectivity is a generalization of ordinary connectivity.
   Instead by a path (sequence of edges), two vertices have to be connected
   by a sequence of short cycles, in which two adjacent cycles have at least
   one common vertex. If all adjacent cycles in the sequence share at least
   one edge, we talk about edge short cycles connectivity.

   It is shown that the short cycles connectivity is an equivalence
   relation on the set of vertices, while the edge short cycles
   connectivity components determine an equivalence relation on the
   set of edges. Efficient algorithms for determining equivalence
   classes are presented.

   Short cycles connectivity can be extended to directed graphs (cyc\-lic
   and transitive connectivity). For further generalization we can also
   consider connectivity by small cliques or other families of graphs.
\end{abstract}


\section{Introduction}

The idea of connectivity by short cycles emerges in different contexts.
In hierarchical decompositions of networks \cite{sym} the long cycles
can be violations of the assumed hierarchical structure -- and related
to general structure these nonhierarhical (cyclic) links can be identified.
The symmetric connectivity from paper \cite{sym} is essentially the
connectivity by 2-cycles. In \cite{triad} we were looking at subgraphs
formed by complete triads -- triangles. Triangular connectivity also
appears to be important in different applications \cite{fri,zip,cur,SW}.

The next stimulus was a reference in Scott \cite{NA} to the early work
of M. Everett on this subject \cite{eva,evb,evc}. It seems that his
ideas can be elaborated to provide a powerful and efficient tool for
analysis of large networks.

In this paper we first present connectivity by cycles of length 3
-- triangular connectivity. Afterward we generalize the results to
connectivity by cycles of length at most $k$, and at the end we
propose further generalizations.

The theorems \ref{equiv.K3}, \ref{max.K3}, \ref{net.K3}, \ref{equiv.L3},
\ref{theorems.3}, \ref{weakstrong.3} are generalized by theorems
\ref{equiv.Kk}, \ref{max.Kk}, \ref{net.Kk}, \ref{equiv.Lk}, \ref{theorems.k}
and \ref{weakstrong.k}. Therefore they are stated without proofs.


\section{Triangular connectivity}

\subsection{Undirected graphs}

Let $\relation{K}$ denotes the \define{connectivity} relation and
$\relation{B}$ denotes the \define{biconnectivity} relation in a
given undirected graph $\graph{G}=(\vertices{V},\edges{E})$.
Let $n=|\vertices{V}|$ denotes the number of vertices and let
$m=|\edges{E}|$ denotes the number of edges.

Vertex $u\in\vertices{V}$ is in relation $\relation{K}$ with vertex
$v\in\vertices{V}$, $u \relation{K} v$, iff $u=v$ or there exists a
path in $\graph{G}$ from $u$ to $v$.

Vertex $u\in\vertices{V}$ is in relation $\relation{B}$ with vertex
$v\in\vertices{V}$, $u \relation{B} v$, iff $u=v$ or there exists a
cycle in $\graph{G}$ containing $u$ and $v$.

We call a \define{triangle} a subgraph isomorphic to a 3-cycle $C_3$.
A subgraph $\graph{H}$ of $\graph{G}$ is \define{triangular}, if
each its vertex and each its edge belong to at least one triangle
in $\graph{H}$.

\begin{definition}
   A sequence $(\graph{T}_1, \graph{T}_2, \ldots, \graph{T}_s)$ of triangles
   of $\graph{G}$ (\define{vertex}) \define{triangularly connects} vertex
   $u\in\vertices{V}$ with vertex $v\in\vertices{V}$, iff
   \begin{enumerate}
      \item $u\in\vertices{V}(\graph{T}_1)$,
      \item $v\in\vertices{V}(\graph{T}_s)$, and
      \item $\vertices{V}(\graph{T}_{i-1})\cap\vertices{V}(\graph{T}_i)\ne\emptyset$
            \quad for \ \ $\between{i}{2}{s}$.
   \end{enumerate}
   Such a sequence is called a (\define{vertex}) \define{triangular chain},
   see Figure \ref{K3}.
\end{definition}

\slika{}{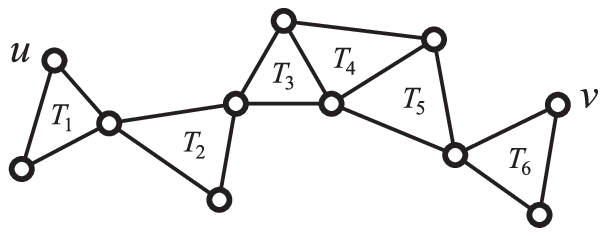}{Triangular chain}

\begin{definition}
   Vertex $u\in\vertices{V}$ is (\define{vertex}) \define{triangularly connected}
   with vertex $v\in\vertices{V}$, $u \relation{K}_3 v$, iff $u=v$ or there
   exists a (vertex) triangular chain that (vertex) triangularly connects
   vertex $u$ with vertex $v$.
\end{definition}

\begin{theorem}
\label{equiv.K3}
   The relation $\relation{K}_3$ is an equivalence relation on the set of vertices
   $\vertices{V}$.
\end{theorem}

A triangular connectivity component is \define{trivial} iff it consists
of a single vertex.

\begin{theorem}
\label{max.K3}
   The sets of vertices of maximal connected triangular subgraphs are
   exactly nontrivial (vertex) triangular connectivity components.
\end{theorem}

But subgraphs, induced by nontrivial (vertex) triangular connectivity components
are not necessary triangular subgraphs and therefore they are not maximal connected
triangular subgraphs. We can see this from example in Figure \ref{tri}, where all
vertices are in the same triangular connectivity component, but the graph is
not triangular because of edge $e$, which does not belong to a triangle.

\slika{}{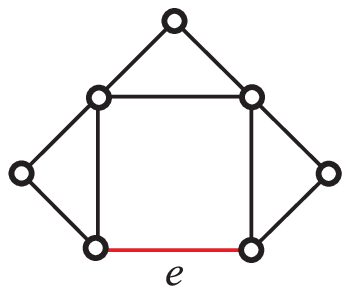}{This graph is not triangular}

An algorithm for determining the relation $\relation{K}_3$ is simple,
see Algorithm \ref{alg:K3}. It partitions the set of vertices into $k$
sets (equivalence classes) labeled $\vertices{C}_1$, $\vertices{C}_2$,
\ldots, $\vertices{C}_k$.

First we choose any vertex $u\in\vertices{V}$ and put it into a new set,
which at the end will become one of the equivalence classes. Then we add
to it all vertices, which can be reached from vertex $u$ by triangles.
We repeat this procedure until all vertices are assigned to equivalence
classes.

\begin{algorithm}[!ht]
\caption{Equivalence classes of the relation $\relation{K}_3$}
\label{alg:K3}
   $k := 0$ \;
   \WhileDo {$\vertices{V}\ne\emptyset$} {
      choose $u\in\vertices{V}$ \;
      $k := k + 1$ \;
      $\vertices{C}_k := \emptyset$ \;
      $\edges{L} := \set{u}$ \;
      \WhileDo {$\edges{L}\ne\emptyset$} {
         choose $u\in\edges{L}$ \;
         $\edges{C}_k := \edges{C}_k \cup \set{u}$ \;
         \ForEachDo {$v\in N(u)$} {
            $\vertices{N} := N(u) \cap N(v)$ \;
            \lIf {$\vertices{N}\ne\emptyset$} {$\edges{L} := \edges{L} \cup \vertices{N} \cup \set{v}$ }
         }
         $\vertices{V} := \vertices{V} \setminus \set{u}$ \;
         $\edges{L} := \edges{L} \setminus \set{u}$ \;
      }
   }
\end{algorithm}

\noindent
$N(u) = \set{v\in\vertices{V}: \edge{u}{v}\in\edges{E}}$ denotes the set of all neighbors
of vertex $u$. If the sets of neighbours are ordered we can use merging to compute
$N(u) \cap N(v)$ in $\OO(\Delta)$, $\Delta$ is the maximum degree of $\graph{G}$.
In this case the time complexity of this algorithm is $\OO(\Delta m)$. We have to
assign each vertex to corresponding equivalence class. To assign vertex $u$, we
have to visit all its neighbors and for each neighbor $v$ we have to find
intersection of $N(u)$ and $N(v)$.

\begin{definition}
   The \define{triangular network} $\network{N}_3(\graph{G})=(\vertices{V},\edges{E}_3,w_3)$
   determined by graph $\graph{G} = (\vertices{V},\edges{E})$ is a subgraph $\graph{G}_3 =
   (\vertices{V},\edges{E}_3)$ of $\graph{G}$ which edges are defined by: $e\in\edges{E}_3$,
   iff $e\in\edges{E}$ and $e$ belongs to a triangle. For edge $e\in\edges{E}_3$ its weight
   $w_3(e)$ equals to the number of different triangles in $\graph{G}$ to which $e$ belongs.
\end{definition}

\begin{theorem}
\label{net.K3}
   $$ \relation{K}_3(\graph{G}) = \relation{K}(\graph{G}_3) $$
\end{theorem}

\noindent
An algorithm for determining $\edges{E}_3$ and $w_3$ is simple, see Algorithm
\ref{alg:E3} and Figure \ref{nn}. If the sets of neighbors are ordered the time
complexity of computing $w_3(e)$ is $\OO(\Delta)$ and the total time complexity
of the algorithm is $\OO(\Delta m)$.

\begin{algorithm}[htbp]
\caption{Triangular network construction}
\label{alg:E3}
   $\edges{E}_3 := \emptyset$ \;
   \ForEachDo {$e\edge{u}{v} \in \edges{E}$} {
      $w_3(e) := |N(u) \cap N(v)|$ \;
      \lIf {$w_3(e)>0$} {$\edges{E}_3 := \edges{E}_3 \cup \set{e}$}
   }
\end{algorithm}

\slika{}{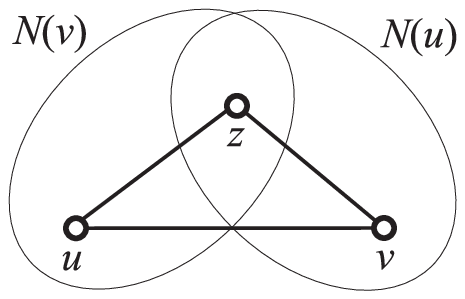}{$w_3(e) := |N(u) \cap N(v)|$}

With $t(v)$ we denote the number of different triangles of $\graph{G}$
that contain vertex $v$. It is easy to verify the following relation
between $t$ and $w$.

\begin{theorem}
   \quad  $\displaystyle{2 t(v) = \sum_{e:e(v:u)} w_3(e)}$
\end{theorem}

\begin{definition}
   A sequence $(\graph{T}_1, \graph{T}_2, \ldots, \graph{T}_s)$ of triangles
   of $\graph{G}$ \define{edge triangularly connects} vertex
   $u\in\vertices{V}$ with vertex $v\in\vertices{V}$, iff
   \begin{enumerate}
      \item $u\in\vertices{V}(\graph{T}_1)$,
      \item $v\in\vertices{V}(\graph{T}_s)$, and
      \item $\edges{E}(\graph{T}_{i-1})\cap\edges{E}(\graph{T}_i)\ne\emptyset$
            \quad for \ \ $\between{i}{2}{s}$.
   \end{enumerate}
   Such a sequence is called an \define{edge triangular chain}, see Figure \ref{L3}.
\end{definition}

\slika{}{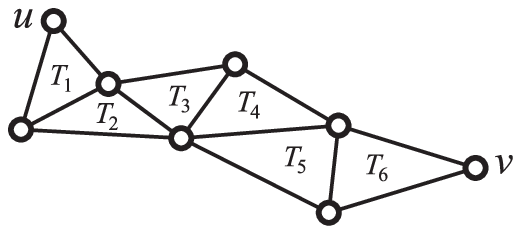}{Edge triangular chain}

\begin{definition}
   Vertex $u\in\vertices{V}$ is \define{edge triangularly connected} with vertex
   $v\in\vertices{V}$, $u \relation{L}_3 v$, iff $u=v$ or there exists an edge
   triangular chain that edge triangularly connects vertex $u$ with vertex $v$.
\end{definition}

In the biconnected graph in Figure \ref{bico} the vertices $u$ in $v$
are edge triangularly connected, while the vertices $x$ and $z$ are not.
The relation $\relation{L}_3$ is not transitive: $x \relation{L}_3 v$,
$v \relation{L}_3 z$, but not $x \relation{L}_3 z$.

\slika{}{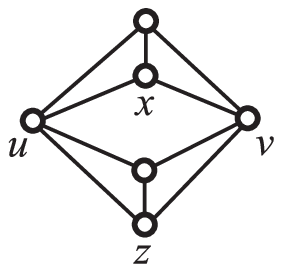}{Biconnected triangular graph}

\begin{theorem}
\label{equiv.L3}
   The relation $\relation{L}_3$ determines an equivalence relation
   on the set of edges $\edges{E}$.
\end{theorem}

An algorithm for determining the relation $\relation{L}_3$ is simple,
see Algorithm \ref{alg:L3}. It partitions the set of edges into $k$
sets (equivalence classes of the relation on $\edges{E}$) labeled
$\vertices{C}_1$, $\vertices{C}_2$, \ldots, $\vertices{C}_k$. Vertex
$u$ is in relation $\relation{L}_3$ with vertex $v$, if both vertices
are end-points of an edge from the same class.

$$
   u\relation{L}_3 v \Leftrightarrow
   \exists i\ \exists e,f\in\vertices{C}_i:
   u\in\vertices{V}(e) \land v \in\vertices{V}(f)
$$

\noindent
Here $\vertices{V}(e)$ denotes the set of end-points of edge $e$.

\begin{algorithm}[htbp]
\caption{Equivalence classes of the relation on $\edges{E}$}
\label{alg:L3}
   $k := 0$ \;
   \WhileDo {$\edges{E}\ne\emptyset$} {
      choose $e\in\edges{E}$ \;
      $k := k + 1$ \;
      $\edges{C}_k := \emptyset$ \;
      $\edges{L} := \set{e}$ \;
      \WhileDo {$\edges{L}\ne\emptyset$} {
         choose $e\edge{u}{v}\in\edges{L}$ \;
         $\edges{C}_k := \edges{C}_k \cup \set{e}$ \;
         $\edges{E} := \edges{E} \setminus \set{e}$ \;
         $\vertices{N} := N(u) \cap N(v)$ \;
         $\edges{L} := \edges{L} \cup \set{\edge{u}{w}, w\in\vertices{N}} \cup \set{\edge{v}{w}, w\in\vertices{N}}$ \;
         $\edges{L} := \edges{L} \setminus \set{e}$ \;
      }
   }
\end{algorithm}

In each iteration of the inner loop we move one edge from $\edges{E}$
into $\edges{C}_k$. So the inner loop repeats $m$-times. Each
assignment or comparison takes constant time, except the statement
where the intersection of two neibourhoods is determined. If the
sets of neighbours are ordered, this statement has time complexity
of $\OO(\Delta)$, so the total time complexity of the algorithm is
$\OO(\Delta m)$.

Note, that in the inner loop the edge $e$ is actually removed from
$\edges{E}$, so the neighborhoods of vertices are dynamical -- they
depend on the current set of edges $\edges{E}$. This means, that after
the edge is removed from $\edges{E}$ (and from $\edges{L}$), it can
not appear in $\edges{L}$ again.

\begin{definition}
   Let $\relation{B}_3 = \relation{B} \cap \relation{K}_3$.
\end{definition}

\begin{theorem}
\label{theorems.3}
   In graph $\graph{G}$ hold:
   \begin{tabbing}
      xxx\=xxxx\=xxxxxxxxxxxxxxxxxxxxxxx\=xxxx\=xxx\kill
      \> a. \> $ \relation{B} \subseteq \relation{K} $
      \> d. \> $ \relation{B}_3 \subseteq \relation{B} $ \\
      \> b. \> $ \relation{K}_3 \subseteq \relation{K} $
      \> e. \> $ \relation{B}_3 \subseteq \relation{K}_3 $ \\
      \> c. \> $ \relation{L}_3 \subseteq \relation{B}_3 $
   \end{tabbing}
\end{theorem}


\subsection{Directed graphs}

If the graph $\graph{G}$ is mixed we replace edges with pairs of opposite arcs.
In the following let $\graph{G}=(\vertices{V},\edges{A})$ be a simple directed
graph without loops.

For a selected arc $a(u,v)\in\edges{A}$ there are four different types of
directed triangles: \textbf{cyc}lic, \textbf{tra}nsitive, \textbf{in}put
and \textbf{out}put.

\slike{
   \podslika[width=20mm]{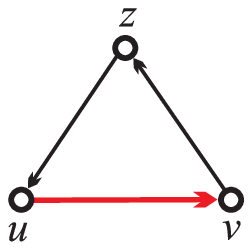}{cyc}
   \podslika[width=20mm]{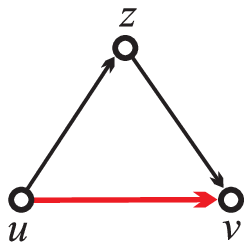}{tra}
   \podslika[width=20mm]{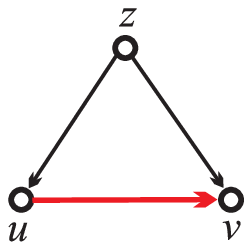}{in}
   \podslika[width=20mm]{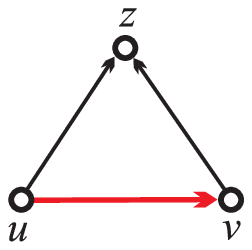}{out}
}{ctio}{Types of directed triangles}

For cyclic triangles we define (similarly as for undirected graphs):

\begin{itemize}
   \item [$\relation{C}_3$] - cyclic triangular connectivity,
   \item [$\relation{D}_3$] - arc cyclic triangular connectivity,
\end{itemize}

\noindent
and the corresponding networks $\network{N}_{cyc}$, $\network{N}_{tra}$,
$\network{N}_{in}$ and $\network{N}_{out}$. The algorithms for determining
relations $\relation{C}_3$ and $\relation{D}_3$ and networks
$\network{N}_{cyc}$, $\network{N}_{tra}$, $\network{N}_{in}$,
$\network{N}_{out}$ are similar to the algorithms for undirected
graphs and have the same complexities.

\begin{theorem}
\label{weakstrong.3}
   A weakly connected cyclic triangular graph is also strongly connected.
\end{theorem}

For $\relation{C}_3$ and $\relation{D}_3$ similar theorems hold as for
$\relation{K}_3$ and $\relation{L}_3$. Besides these two connectivities
there is another possibility. Both graphs in Figure \ref{strcy} are (weakly)
triangular. The left side graph is also cyclic triangularly connected,
but the right side graph is not. This leads to the following definition.

\slike{
   \podslika[width=22mm]{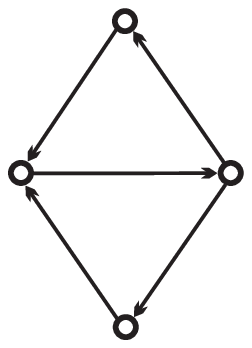}{cyclic}
   \podslika[width=22mm]{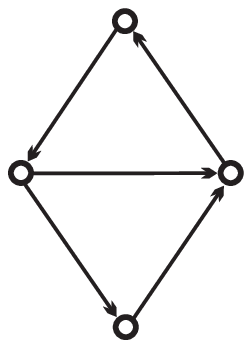}{noncyclic}
}{strcy}{Strongly triangularly connected graphs}

The vertices $u,v\in\vertices{V}$ are (\define{vertex}) \define{strongly
triangularly connected}, $u \relation{S}_3 v$, iff $u=v$ or there exists
strongly connected triangular subgraph that contains $u$ and $v$.


\subsection{Transitivity}

Let $\relation{R}$ denotes the \define{reachability} relation
in a given directed graph $\graph{G}=(\vertices{V},\edges{A})$.
Vertex $v$ is \define{reachable} from vertex $u$, $u \relation{R} v$,
iff $u=v$ or there exists a walk from $u$ to $v$.

\begin{theorem}
   If we remove from a graph $\graph{G}=(\vertices{V},\edges{A})$ all
   (or some) arcs belonging to a triangularly transitive path $\pi$
   (all arcs of $\pi$ are transitive) the reachability relation does not change:
   $\relation{R}(\graph{G}) = \relation{R}(\graph{G}\setminus \edges{A}(\pi))$.
\end{theorem}

\begin{proof}
   Because the graph $\graph{G}\setminus\edges{A}(\pi)$ is a subgraph of $\graph{G}$,
   it is obvious that $\relation{R}(\graph{G}\setminus\edges{A}(\pi)) \subseteq
   \relation{R}(\graph{G})$. Let $a$ be any arc on the transitive path $\pi$.
   Because of its transitivity, its terminal vertex is also reachable from its
   initial vertex by two supporting arcs. We have only to check, that none of them
   is a part of the path $\pi$, so it can not be deleted. Because the arc $a$
   and its supporting arc have a common vertex, the only way to be on the same
   path is to be subsequent arcs. But this is impossible because of their
   directions. So also the opposite is true: $\relation{R}(\graph{G})\subseteq
   \relation{R}(\graph{G}\setminus\edges{A}(\pi))$.
\end{proof}

\bigskip
But, we cannot remove all transitive arcs. The counter-example is presented in
Figure \ref{reachability}, where we have a directed 6-cycle, which vertices are
connected by arcs with additional vertex in its center. The central vertex is
reachable from anywhere. All the arcs from the cycle to the central vertex are transitive.
If we remove them all, the central vertex is not reachable any more.

\slika{width=30mm}{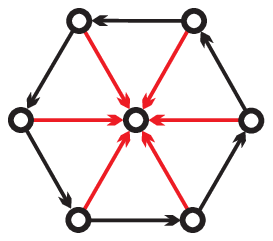}{Graph in which all transitive arcs can not be removed}


\section{$k$-gonal connectivity}

\subsection{Undirected graphs}

We call a \define{$k$-gone} a subgraph isomorphic to a $k$-cycle $C_k$ and a
\define{$(k)$-gone} a subgraph isomorphic to $C_s$ for some $s$, $3\le s\le k$.
A subgraph $\graph{H}$ of $\graph{G}$ is \define{$k$-gonal}, if
each its vertex and each its edge belong to at least one $(k)$-gone
in $\graph{H}$.

\begin{definition}
   A sequence $(\graph{C}_1, \graph{C}_2, \ldots, \graph{C}_s)$ of $(k)$-gones
   of $\graph{G}$ (\define{vertex}) \define{$k$-go\-nally connects}
   vertex $u\in\vertices{V}$ with vertex $v\in\vertices{V}$, iff
   \begin{enumerate}
      \item $u\in\vertices{V}(\graph{C}_1)$,
      \item $v\in\vertices{V}(\graph{C}_s)$, and
      \item $\vertices{V}(\graph{C}_{i-1})\cap\vertices{V}(\graph{C}_i)\ne\emptyset$
         \quad for \ \ $\between{i}{2}{s}$.
   \end{enumerate}
   Such a sequence is called a (\define{vertex}) \define{$k$-gonal chain},
   see Figure \ref{Kk}.
\end{definition}

\slika{}{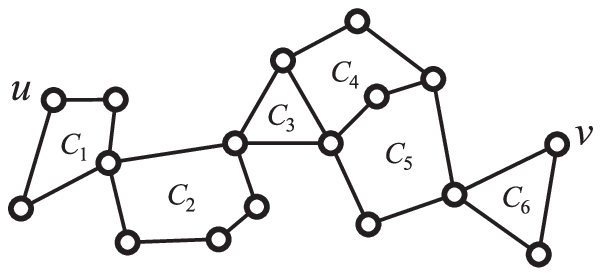}{$k$-gonal chain}

\begin{definition}
   Vertex $u\in\vertices{V}$ is (\define{vertex}) \define{$k$-gonally connected}
   with vertex $v\in\vertices{V}$, $u \relation{K}_k v$, iff $u=v$ or there exists
   a (vertex) $k$-gonal chain that (vertex) $k$-gonally connects vertex $u$ with
   vertex $v$.
\end{definition}

\begin{theorem}
   \label{equiv.Kk}
   The relation $\relation{K}_k$ is an equivalence relation on the set of vertices
   $\vertices{V}$.
\end{theorem}

\begin{proof}
   Reflexivity follows directly from the definition of the relation $\relation{K}_k$.

   Since the reverse of a $k$-gonal chain from $u$ to $v$ is a $k$-gonal
   chain from $v$ to $u$, the relation $\relation{K}_k$ is symmetric.

   Transitivity. Let $u$, $v$ and $z$ be such vertices, that $u\relation{K}_k v$
   and $v\relation{K}_k z$. If these vertices are not pairwise different, the
   transitivity condition is trivially true. Assume now that they are pairwise
   different. Because of $u\relation{K}_k v$ and $v\relation{K}_k z$ there exist
   (vertex) $k$-gonal chains from $u$ to $v$ and from $v$ to $z$. Their
   concatenation is a (vertex) $k$-gonal chain from $u$ to $z$. Therefore
   also $u\relation{K}_k z$.
\end{proof}

\begin{theorem}
\label{max.Kk}
   The sets of vertices of maximal connected $k$-gonal subgraphs are
   exactly nontrivial (vertex) $k$-gonal connectivity components.
\end{theorem}

\begin{proof}
   Let $u$ and $v$ be any vertices belonging to a connected $k$-gonal subgraph.
   If $u=v$, it is obvious that $u\relation{K}_k v$. Otherwise there exists a
   path $\pi = u, e_1, z_1, e_2, z_2, e_3, z_3, \ldots, e_s, v$ from $u$ to $v$.
   Because the subgraph is $k$-gonal, each edge $e_i$ on this path belongs to at
   least one $(k)$-gone $\graph{C}_i$ in this subgraph. For the obtained $k$-gonal
   chain $(\graph{C}_1, \graph{C}_2, \ldots, \graph{C}_s)$ it holds:
   \begin{itemize}
      \item $e_i\in\edges{E}(\graph{C}_i)$, $\between{i}{1}{s}$
      \item $u\in\vertices{V}(\graph{C}_1)$, $v\in\vertices{V}(\graph{C}_s)$
      \item $z_{i-1}\in\vertices{V}(\graph{C}_{i-1})\cap \vertices{V}(\graph{C}_i)$,
            $\between{i}{2}{s}$
   \end{itemize}
   Therefore $u\relation{K}_k v$. So all vertices of any (also maximal)
   connected $k$-gonal subgraph belong to the same component of the
   relation $\relation{K}_k$.

   Now let $u$ and $v$ be two different vertices of a nontrivial
   $\relation{K}_k$-component $\vertices{C}\subseteq\vertices{V}$.
   Because $u$ is in relation $\relation{K}_k$ with $v$, there exists
   a $k$-gonal chain from $u$ to $v$. It is obvious that all vertices
   of a $k$-gonal chain belong to the same maximal connected $k$-gonal
   subgraph, so also $u$ and $v$. But $u$ and $v$ were any two different
   vertices of $\vertices{C}$, so all vertices of a nontrivial $k$-gonal
   connectivity component belong to the same maximal connected
   $k$-gonal subgraph.
\end{proof}

\bigskip
But, as shown in Figure \ref{tri}, subgraphs induced by nontrivial (vertex)
$k$-gonal connectivity components are not necessary $k$-gonal subgraphs and
therefore they are not maximal connected $k$-gonal subgraphs.

\begin{definition}
   The \define{$k$-gonal network} $\network{N}_k(\graph{G})=(\vertices{V},\edges{E}_k,w_k)$
   determined by graph $\graph{G} = (\vertices{V},\edges{E})$ is a subgraph $\graph{G}_k =
   (\vertices{V},\edges{E}_k)$ of $\graph{G}$ which edges are defined by: $e\in\edges{E}_k$,
   iff $e\in\edges{E}$ and $e$ belongs to a $(k)$-gone. For an edge $e\in\edges{E}_k$ the weight
   $w_k(e)$ equals to the number of different $(k)$-gones in $\graph{G}$ to which $e$ belongs.
\end{definition}

\begin{theorem}
   \label{net.Kk}
   $$ \relation{K}_k(\graph{G}) = \relation{K}(\graph{G}_k) $$
\end{theorem}

\begin{proof}
   Let $u\relation{K}_k v$ holds in graph $\graph{G}$. If $u=v$, it is also
   true that $u\relation{K} v$ in graph $\graph{G}_k$. If the vertices
   are different, there exists (vertex) $k$-gonal chain in $\graph{G}$
   from $u$ to $v$. Each edge in this chain belongs to at least one
   $(k)$-gone, so the whole chain is in $\graph{G}_k$. So $u$ and $v$
   are connected in $\graph{G}_k$ or with other words $u\relation{K} v$
   in $\graph{G}_k$.

   Let $u\relation{K} v$ holds in graph $\graph{G}_k$. Then in graph
   $\graph{G}_k$ exists a path from $u$ to $v$. Because $\graph{G}_k$
   is $k$-gonal, each edge on this path belongs to at least one $(k)$-gone,
   so we can construct a $k$-gonal chain from $u$ to $v$ in $\graph{G}_k$.
   Because $\graph{G}_k$ is subgraph of $\graph{G}$, this chain is also
   in $\graph{G}$, which means that $u\relation{K}_k v$ in graph $\graph{G}$.
\end{proof}

\bigskip
To determine the equivalence classes of the relation $\relation{K}_k$,
we can first determine its $k$-gonal subgraph $\graph{G}_k$ and find the
connected components of it.

To compute the weight of edge $e$ we have to count to how many $(k)$-gones
it belongs. We are still working on development of an efficient algorithm for
this task.


The weights $w_k$ can be used to identify dense parts of a given network.
For example, for a selected edge $e$ in $r$-clique we can count, to how many $k$-gones it belongs.
The end-points of $e$ are the first two vertices of the $k$-gone. There are $r-2$
ways to choose the third vertex, then $r-3$ ways to choose the fourth vertex, ...,
and $r-k+1$ ways to choose the last vertex of the $k$-gone (which is connected to
the first one). So we have $(r-2)(r-3)\cdots(r-k+1)$ different $k$-gones and
$\sum_{i=3}^k (r-2)(r-3)\cdots(r-i+1)$ different $(k)$-gones. It follows that
each edge $e$ of $r$-clique in $k$-gonal network has weight $w_k(e)$ at least
$\sum_{i=3}^k (r-2)(r-3)\cdots(r-i+1)$

$$ w_k(e) \ge \sum_{i=3}^k (r-2)(r-3)\cdots(r-i+1) $$

The \define{Everett's $k$-decomposition} of a given undirected graph $\graph{G} =
(\vertices{V},\edges{E})$ is a partition $\set{\vertices{C}_1, ..., \vertices{C}_p,
\vertices{B}_1, ..., \vertices{B}_q}$ of the set of vertices $\vertices{V}$,
where $\vertices{C}_i$ are the $k$-gonally connected components and $\vertices{B}_j$
are \define{bridges} -- the connected components of the $\vertices{V} \setminus \cup
\vertices{C}_i$.

A procedure for determining Everett's decomposition is as follows: First
determine the $k$-gonal subgraph $\graph{G}_k$. Its connected components
$\set{\vertices{C}_i}$ are by Theorem \ref{net.Kk} just the $k$-gonally
connected components. Finally in the graph $\graph{G} | \vertices{V}
\setminus \cup \vertices{C}_i$ determine the connected components --
bridges $\set{\vertices{B}_i}$.

\begin{definition}
   A sequence $(\graph{C}_1, \graph{C}_2, \ldots, \graph{C}_s)$ of $(k)$-gones
   of $\graph{G}$ \define{edge $k$-gonally connects} vertex $u\in\vertices{V}$
   with vertex $v\in\vertices{V}$, iff
   \begin{enumerate}
      \item $u\in\vertices{V}(\graph{C}_1)$,
      \item $v\in\vertices{V}(\graph{C}_s)$, and
      \item $\edges{E}(\graph{C}_{i-1})\cap\edges{E}(\graph{C}_i)\ne\emptyset$
         \quad for \ \ $\between{i}{2}{s}$.
   \end{enumerate}
   Such a sequence is called an \define{edge $k$-gonal chain}, see Figure \ref{Lk}.
\end{definition}

\slika{}{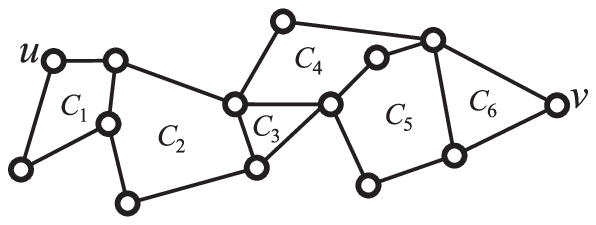}{Edge $k$-gonal chain}

\begin{definition}
   Vertex $u\in\vertices{V}$ is \define{edge $k$-gonally connected} with vertex
   $v\in\vertices{V}$, $u \relation{L}_k v$, iff $u=v$ or there exists an edge
   $k$-gonal chain that edge $k$-gonally connects vertex $u$ with vertex $v$.
\end{definition}

\begin{theorem}
   \label{equiv.Lk}
   The relation $\relation{L}_k$ determines an equivalence relation
   on the set of edges $\edges{E}$.
\end{theorem}

\begin{proof}
   Let the relation $\sim$ on $\edges{E}$ be defined as: $e\sim f$, iff $e=f$
   or there exists an edge $k$-gonal chain $(\graph{C}_1, \graph{C}_2, \ldots,
   \graph{C}_s)$, where $e\in\edges{E}(\graph{C}_1)$ and $f\in\edges{E}(\graph{C}_s)$.

   Reflexivity of $\sim$ follows from its definition.

   The symmetry is simple too. Let be $e\sim f$. Then there exists an
   edge $k$-gonal chain 'from' $e$ 'to' $f$. Its reverse is an edge $k$-gonal
   chain 'from' $f$ 'to' $e$, so $f\sim e$.

   And transitivity. Let $e$, $f$ and $g$ be such edges, that $e\sim f$
   and $f\sim g$. There exist an edge $k$-gonal chain from $e$ to $f$ and
   an edge $k$-gonal chain from $f$ to $g$. The concatenation of these two
   chains is an edge $k$-gonal chain from $e$ to $g$ (the $(k)$-gones in
   the contact of the chains both contain the edge $f$, so their
   intersection is not empty). Therefore also $e\sim g$.
\end{proof}



\begin{definition}
   Let $\relation{B}_k = \relation{B} \cap \relation{K}_k$.
\end{definition}

\begin{theorem}
   \label{theorems.k}
   In graph $\graph{G}$ hold:
   \begin{tabbing}
      xxx\=xxxx\=xxxxxxxxxxxxxxxxxxxxxxx\=xxxx\=xxx\kill
      \> a. \> $ \relation{B} \subseteq \relation{K} $
      \> d. \> $ \relation{B}_k \subseteq \relation{B} $ \\
      \> b. \> $ \relation{K}_k \subseteq \relation{K} $
      \> e. \> $ \relation{B}_k \subseteq \relation{K}_k $ \\
      \> c. \> $ \relation{L}_k \subseteq \relation{B}_k $
   \end{tabbing}
   and for $i < j$ also:
   \begin{tabbing}
      xxx\=xxxx\=xxxxxxxxxxxxxxxxxxxxxxx\=xxxx\=xxx\kill
      \> f. \> $ \mathbf{K}_i \subseteq \mathbf{K}_j $
      \> h. \> $ \mathbf{B}_i \subseteq \mathbf{B}_j $ \\
      \> g. \> $ \mathbf{L}_i \subseteq \mathbf{L}_j $
   \end{tabbing}
\end{theorem}

\begin{proof}
   \begin{enumerate}
      \item [$a$.] Evident from definitions.

      \item [$b$.] Let $u$ and $v$ be such vertices, that $u\relation{K}_k v$.
      If $u = v$, it is also $u\relation{K}v$ by the definition. Otherwise there
      exists $k$-gonal chain from $u$ to $v$. Therefore there also exists a path
      from $u$ to $v$, which means that $u\relation{K}v$.


      \item [$c$.] Let $u$ and $v$ be such vertices, that $u\relation{L}_k v$.
      If $u=v$, it is also $u\relation{B} v$ and $u\relation{K}_k v$ by definition,
      from which it follows that $u\relation{B}_k v$. If the vertices are different,
      there exists an edge $k$-gonal chain from $u$ to $v$. But since each edge $k$-gonal
      chain is also a vertex $k$-gonal chain (if two $(k)$-gones have a common edge,
      they also have a common vertex), $u\relation{K}_k v$ holds. Subgraph in the form
      of an edge $k$-gonal chain is biconnected \cite{GT}, $u\relation{B} v$. Both
      results together give us $u\relation{B}_k v$.

      \item [$d$.] Follows from the definition of the relation $\relation{B}_k$.
      \item [$e$.] Follows from the definition of the relation $\relation{B}_k$.

      \item [$f$.] Let $u$ and $v$ be such vertices, that $u\relation{K}_i v$.
      If $u=v$, it is also $u\relation{K}_j v$ by the definition. Otherwise
      there exists $i$-gonal chain from $u$ to $v$, where none of $(i)$-gones
      has length greater than $i$. The same chain is also a $j$-gonal chain
      from $u$ to $v$ -- therefore $u\relation{K}_j v$.

      \item [$g$.] Let $u$ and $v$ be such vertices, that $u\relation{L}_i v$.
      If $u=v$, it is also $u\relation{L}_j v$ by the definition. Otherwise
      there exists edge $i$-gonal chain from $u$ to $v$, where none of $(i)$-gones
      has length greater than $i$. The same chain is also an edge $j$-gonal chain
      from $u$ to $v$ -- therefore $u\relation{L}_j v$.

      \item [$h$.] Follows from the definition of the relation $\relation{B}_k$
      and item $f$ of this theorem.
   \end{enumerate}
\end{proof}

The relationships from theorem \ref{theorems.k} can be presented by a diagram:

$$
   \begin{array}{ccccc}
      & & \relation{B} & \subseteq & \relation{K} \\
      \rotatebox{90}{$\subseteq...$} & & \rotatebox{90}{$\subseteq...\subseteq$} & & \rotatebox{90}{$\subseteq...\subseteq$} \\
      \relation{L}_k & \subseteq & \relation{B}_k & \subseteq & \relation{K}_k \\
      \rotatebox{90}{$\subseteq$} & & \rotatebox{90}{$\subseteq$} & & \rotatebox{90}{$\subseteq$} \\
      \relation{L}_{k-1} & \subseteq & \relation{B}_{k-1} & \subseteq & \relation{K}_{k-1} \\
      \rotatebox{90}{$...\subseteq$} & & \rotatebox{90}{$...\subseteq$} & & \rotatebox{90}{$...\subseteq$} \\
   \end{array}
$$


\subsection{Directed graphs}

We shall give a special attention to two special types of Everett's semicycles
\cite{eva,evb}, see Figure \ref{arcy}, related to selected arc $a(u,v)\in A$:
\textbf{cycles} (arc with a feed-back path) and \textbf{transitive semicycles}
(arc with a reinforcement path) of length at most $k$.

\slike{
   \podslika{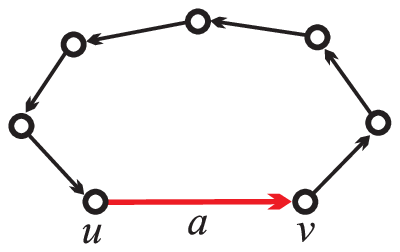}{cyclic \\ feed-back}
   \podslika{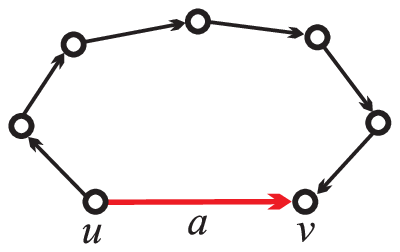}{transitive \\ reinforcement}
}{arcy}{Cycles on an arc}

For cyclic $(k)$-gones we define (similarly as for undirected graphs):

\begin{definition}
   A sequence $(\graph{C}_1, \graph{C}_2, \ldots, \graph{C}_s)$ of cycles
   of length at most $k$ and at least 2 of $\graph{G}$ (\define{vertex})
   \define{cyclic $k$-gonally connects} vertex $u\in\vertices{V}$ with vertex
   $v\in\vertices{V}$, iff
   \begin{enumerate}
      \item $u\in\vertices{V}(\graph{C}_1)$,
      \item $v\in\vertices{V}(\graph{C}_s)$, and
      \item $\vertices{V}(\graph{C}_{i-1})\cap\vertices{V}(\graph{C}_i)\ne\emptyset$
         \quad for \ \ $\between{i}{2}{s}$.
   \end{enumerate}
   Such a sequence is called a (\define{vertex}) \define{cyclic $k$-gonal chain}.
\end{definition}

\begin{definition}
   Vertex $u\in\vertices{V}$ is (\define{vertex}) \define{cyclic $k$-gonally connected}
   with vertex $v\in\vertices{V}$, $u \relation{C}_k v$, iff $u=v$ or there exists
   a (vertex) cyclic $k$-gonal chain that (vertex) cyclic $k$-gonally connects vertex
   $u$ with vertex $v$.
\end{definition}

\begin{definition}
   A sequence $(\graph{C}_1, \graph{C}_2, \ldots, \graph{C}_s)$ of cycles of
   length at most $k$ and at least 2 of $\graph{G}$ \define{arc cyclic $k$-gonally
   connects} vertex $u\in\vertices{V}$ with vertex $v\in\vertices{V}$, iff
   \begin{enumerate}
      \item $u\in\vertices{V}(\graph{C}_1)$,
      \item $v\in\vertices{V}(\graph{C}_s)$, and
      \item $\edges{A}(\graph{C}_{i-1})\cap\edges{A}(\graph{C}_i)\ne\emptyset$
         \quad for \ \ $\between{i}{2}{s}$.
   \end{enumerate}
   Such a sequence is called an \define{arc cyclic $k$-gonal chain}.
\end{definition}

\begin{definition}
   Vertex $u\in\vertices{V}$ is \define{arc cyclic $k$-gonally connected}
   with vertex $v\in\vertices{V}$, $u \relation{D}_k v$, iff $u=v$ or there
   exists an arc cyclic $k$-gonal chain that arc cyclic $k$-gonally connects
   vertex $u$ with vertex $v$.
\end{definition}

For $\relation{C}_k$ and $\relation{D}_k$ similar theorems hold as
for $\relation{K}_k$ and $\relation{L}_k$.

\begin{theorem}
\label{weakstrong.k}
   A weakly connected cyclic $k$-gonal graph is also strongly connected.
\end{theorem}

\begin{proof}
   Take any pair of vertices $u$ and $v$. Since $\graph{G}$ is weakly connected
   there exists a semipath connecting $u$ and $v$. Each arc on this semipath
   belongs to at least one $(k)$-cycle. Therefore its end-points are connected
   by a path in opposite direction -- we can construct a walk from $u$ to $v$
   and also a walk from $v$ to $u$.
\end{proof}

\begin{theorem}
   Cyclic $k$-gonal connectivity $\relation{C}_k$ is an equivalence relation
   on the set of vertices $\vertices{V}$.
\end{theorem}

\begin{proof}
   Reflexivity follows directly from the definition of the relation $\relation{C}_k$.

   Since the reverse of a cyclic $k$-gonal chain from $u$ to $v$ is a cyclic
   $k$-gonal chain from $v$ to $u$, the relation $\relation{C}_k$ is symmetric.

   Transitivity. Let $u$, $v$ and $z$ be such vertices, that $u\relation{C}_k v$
   and $v\relation{C}_k z$. If the vertices are not pairwise different, the
   transitivity condition is trivialy true. Assume now that they are pairwise
   different. Because of $u\relation{C}_k v$ and $v\relation{C}_k z$ there
   exist cyclic $k$-gonal chains from $u$ to $v$ and from $v$ to $z$.
   Their concatenation is a cyclic $k$-gonal chain from $u$ to $z$. Therefore
   also $u\relation{C}_k z$.
\end{proof}

\bigskip
An arc is \define{cyclic} iff it belongs to some cycle (of any length) in the
graph $\graph{G} = (\vertices{V},\edges{A})$. The cyclic arcs that do not
belong to some $(k)$-cycle are called \define{$k$-long} (range) arcs.

\begin{theorem}
   If the graph $\graph{G} = (\vertices{V},\edges{A})$ does not contain
   $k$-long arcs then its cyclic $k$-gonal reduction
   $\graph{G}/\relation{C}_k = (\vertices{V}/\relation{C}_k,\edges{A}^*)$,
   where for $X, Y \in \vertices{V}/\relation{C}_k :
   (X,Y) \in \edges{A}^* \Longleftrightarrow \exists u \in X \exists v \in Y:
   (u,v) \in \edges{A}$, is an acyclic graph.
\end{theorem}

\begin{proof}
   Suppose that cyclic $k$-gonal reduction of graph $\graph{G}$ is not acyclic.
   Then it contains a cycle $C^*$, which can be extended to a cycle $C$ of
   graph $\graph{G}$. Let $a^*$ be any arc of $C^*$ and let $a$ be a
   corresponding arc of $C$. Because the end-points of $a^*$ are different,
   the end-points of $a$ belong to two different components of the relation
   $\relation{C}_k$. So $a$ does not belong to any cyclic $(k)$-gone.
   But $a$ is cyclic (it belongs to cycle $C$), so it is a $k$-long arc.
   This is a contradiction. Therefore, the cyclic $k$-gonal reduction of graph
   $\graph{G}$ must be acyclic.
\end{proof}

From this proof we also see how to identify the $k$-long arcs. They are
exactly the arcs that are reduced to cyclic arcs in $\graph{G}/\relation{C}_k$.

\begin{theorem}
   The relation $\relation{D}_k$ determines an equivalence relation on the
   set of arcs $\edges{A}$.
\end{theorem}

\begin{proof}
   Let the relation $\sim$ on $\edges{A}$ be defined as: $e\sim f$, iff $e=f$
   or there exists an arc cyclic $k$-gonal chain $(\graph{C}_1, \graph{C}_2, \ldots,
   \graph{C}_s)$, where $e\in\edges{A}(\graph{C}_1)$ and $f\in\edges{A}(\graph{C}_s)$.

   Reflexivity of $\sim$ follows from its definition.

   The symmetry is simple too. Let be $e\sim f$. Then there exists an
   arc cyclic $k$-gonal chain 'from' $e$ 'to' $f$. Its reverse is an
   arc cyclic $k$-gonal chain 'from' $f$ 'to' $e$, so $f\sim e$.

   And transitivity. Let $e$, $f$ and $g$ be such arcs, that $e\sim f$
   and $f\sim g$. There exist an arc cyclic $k$-gonal chain from $e$ to $f$
   and an arc cyclic $k$-gonal chain from $f$ to $g$. The concatenation of
   these two chains is an arc cyclic $k$-gonal chain from $e$ to $g$
   (the $(k)$-cycles in the contact of the chains both contain the arc $f$,
   so their intersection is not empty). Therefore also $e\sim g$.
\end{proof}

\begin{definition}
   The vertices $u,v\in\vertices{V}$ are (\define{vertex}) \define{strongly
   $k$-gonally connected}, $u\relation{S}_k v$, iff $u=v$ or there exists
   strongly connected $k$-gonal subgraph that contains $u$ and $v$.
\end{definition}

It is easy to see that
$\relation{D}_k \subseteq \relation{C}_k \subseteq \relation{S}_k$.
The relationships between these relations can be presented by a diagram:

$$
   \begin{array}{ccccc}
      & & & & \relation{S} \\
      \rotatebox{90}{$\subseteq...$} & & \rotatebox{90}{$\subseteq...$} & & \rotatebox{90}{$\subseteq...\subseteq$} \\
      \relation{D}_k & \subseteq & \relation{C}_k & \subseteq & \relation{S}_k \\
      \rotatebox{90}{$\subseteq$} & & \rotatebox{90}{$\subseteq$} & & \rotatebox{90}{$\subseteq$} \\
      \relation{D}_{k-1} & \subseteq & \relation{C}_{k-1} & \subseteq & \relation{S}_{k-1} \\
      \rotatebox{90}{$...\subseteq$} & & \rotatebox{90}{$...\subseteq$} & & \rotatebox{90}{$...\subseteq$} \\
   \end{array}
$$

We can define three networks, that can provide us with more
detailed picture about the network structure:

\begin{itemize}
   \item
   \define{Feedback network} $\network{N}_F = (\vertices{V},\edges{A}_F,w_F)$ where
   $w_F(a)$ is the number of different $(k)$-cycles containing the arc $a$.

   \item
   \define{Transitive network} $\network{N}_T = (\vertices{V},\edges{A}_T,w_T)$ where
   $w_T(a)$ is the number of different transitive $(k)$-semicycles containing the arc
   $a$ as the transitive arc (shortcut).

   \item
   \define{Support network} $\network{N}_S = (\vertices{V},\edges{A}_S,w_S)$ where
   $w_S(a)$ is the number of different transitive $(k)$-semicycles containing the arc
   $a$ as a nontransitive arc.
\end{itemize}

\begin{theorem}
   Let $\relation{S}$ be the relation of strong connectivity, $\relation{S}=
   \relation{R}\cap\relation{R}^{-1}$. Then
   $$ \relation{C}_k(\graph{G}) = \relation{S}(\graph{G}_F) $$
\end{theorem}

\begin{proof}
   Let $u\relation{C}_k v$ holds in graph $\graph{G}$. If $u=v$, it is also
   true, that $u\relation{S} v$ in graph $\graph{G}_F$. If the vertices
   are different, there exists cyclic $k$-gonal chain in $\graph{G}$
   from $u$ to $v$. Each arc in this chain belongs to at least one
   $(k)$-cycle, so the whole chain is in $\graph{G}_F$. Vertices
   $u$ and $v$ are mutually reachable by arcs of this chain, so
   $u\relation{S} v$ in $\graph{G}_F$.

   Let $u\relation{S} v$ holds in graph $\graph{G}_F$. Then in graph
   $\graph{G}_F$ exists a walk from $u$ to $v$. Because $\graph{G}_F$
   is cyclic $k$-gonal, each arc on this walk belongs to at least one $(k)$-cycle,
   so we can construct a cyclic $k$-gonal chain from $u$ to $v$ in $\graph{G}_F$.
   Because $\graph{G}_F$ is subgraph of $\graph{G}$, this chain is also
   in $\graph{G}$, which means that $u\relation{C}_k v$ holds in graph $\graph{G}$.
\end{proof}


\subsection{Transitivity}

Let $\relation{T}_k$ denotes the $k$-transitive reachability relation
in a given directed graph $\graph{G}=(\vertices{V},\edges{A})$.

\begin{definition}
   Vertex $v$ is \define{$k$-transitively reachable} from vertex $u$,
   $u \relation{T}_k v$, iff $u=v$ or there exists a walk from $u$
   to $v$ in which each arc is $k$-transitive -- is a base (shortcut
   arc) of some transitive semicycle of length at most $k$.
\end{definition}

The vertices $u$ and $v$ are mutually $k$-transitively reachable,
if vertex $u$ is $k$-transitively reachable from vertex $v$, and
vertex $v$ is $k$-transitively reachable from vertex $u$.
We denote this relation by $\hat{\relation{T}}_k$
$$ u\hat{\relation{T}}_k v \Leftrightarrow u\relation{T}_k v \land v\relation{T}_k u $$

\begin{theorem}
   The relation of mutual $k$-transitive reachability $\hat{\relation{T}}_k =
   \relation{T}_k\cap\relation{T}_k^{-1}$ is an equivalence relation
   on the set of vertices $V$.
\end{theorem}

\begin{proof}
   It is well known that if $\relation{Q}$ is a reflexive and transitive relation
   then $\hat{\relation{Q}} = \relation{Q}\cap\relation{Q}^{-1}$ is an equivalence
   relation. The relation $\relation{T}_k$ is reflexive by definition, so we have
   only to prove that it is transitive.

   Let $u$, $v$ and $w$ be such vertices that $u\relation{T}_k v$ and $v\relation{T}_k w$.
   If these vertices are not pairwise different, the transitivity condition is
   trivialy true. Otherwise there exists a walk from $u$ to $v$ and a walk from
   $v$ to $w$, in which every arc is $k$-transitive. Their concatenation is a walk
   from $u$ to $w$, in which every arc is $k$-transitive, so $u\relation{T}_k w$.
\end{proof}


\section{Further generalizations}

Until now we observed the connectivity by triangles and other short cycles.
Intersections of two adjacent cycles in the coresponding chains contained at
least one vertex (vertex connectivity) or at least one edge/arc (edge/arc
connectivity). This can be generalized to other families of grahps.

\begin{definition}
   Let $\HH$ and $\HH_0$ be two families of graphs.
   A sequence $(\graph{H}_1, \graph{H}_2,$ $\ldots, \graph{H}_s)$ of subgraphs
   of $\graph{G}$ \define{$(\HH,\HH_0)$ connects} vertex $u\in\vertices{V}$
   with vertex $v\in\vertices{V}$, iff
   \begin{enumerate}
      \item $u\in\vertices{V}(\graph{H}_1)$,
      \item $v\in\vertices{V}(\graph{H}_s)$,
      \item $\graph{H}_i\in\HH$ \quad for \ \ $\between{i}{1}{s}$, and
      \item $\graph{H}_{i-1}\cap\graph{H}_i\supseteq\graph{H}\in\HH_0$
         \quad for \ \ $\between{i}{2}{s}$.
   \end{enumerate}
\end{definition}

\begin{example}
   For $r<k$ we can define $(k,r)$-clique connectivity:
   $\HH=\{K_{r+1},K_{r+2},\ldots,K_k\}$, $\HH_0=\{K_r\}$
\end{example}

\bigskip
All the previous types of connectivity are special cases of the
generalized connectivity:
\begin{eqnarray*}
   \relation{K}_3 & = & \mbox{$(3,1)$-clique connectivity} \\
   \relation{L}_3 & = & \mbox{$(3,2)$-clique connectivity} \\
   \relation{K}_k & = & \mbox{$(\{C_3, \ldots, C_k\}, \{K_1\})$ connectivity} \\
   \relation{L}_k & = & \mbox{$(\{C_3, \ldots, C_k\}, \{K_2\})$ connectivity}
\end{eqnarray*}

For the generalized connectivity similar theorems hold as for
triangular and $k$-gonal connectivity.


\section*{Acknowledgements}

This work was supported by the Ministry of Education, Science and Sport of
Slovenia, Project 0512--0101. Special thanks to Martin G. Everett for copies
of his papers on the subject.

The paper was presented at \emph{Fifth Slovenian International Conference
On Graph Theory}, June 22--27, 2003, Bled, Slovenia.



\end{document}